\documentclass[aps,pra,twocolumn,showpacs,superscriptaddress,groupedaddress,amsmath,amssymb,longbibliography]{revtex4-1}
\usepackage{graphicx}
\usepackage{subfigure}
\usepackage{color}
\usepackage{dcolumn}
\newcommand{\be}{\begin{equation}}
\newcommand{\ee}{\end{equation}}
\newcommand{\bea}{\begin{eqnarray}}
\newcommand{\eea}{\end{eqnarray}}
\newcommand{\ba}{\begin{array}}
\newcommand{\ea}{\end{array}}

\newtheorem{thm}{Theorem}

\begin{document}

\title{Quantum code for quantum error characterization}
%\title{Ambiguous stabilizer codes  and quantum noise characterization}
\author{S.  Omkar}  \affiliation{Poornaprajna Institute  of Scientific
  Research,  Sadashivnagar,  Bengaluru-  560080,  India.}   \author{R.
  Srikanth}   \email{srik@poornaprajna.org}  \affiliation{Poornaprajna
  Institute of Scientific  Research, Sadashivnagar, Bengaluru- 560080,
  India.}  \author{Subhashish    Banerjee}
\affiliation{Indian  Institute  of  Technology  Jodhpur,  Rajasthan  -
  342011, India}

\begin{abstract} 
A quantum error correcting code  is a subspace $\mathcal{C}$ such that
allowed errors acting on any  state in $\mathcal{C}$ can be corrected.
A  quantum code  for which  state recovery  is only  required up  to a
logical rotation  within $\mathcal{C}$, can  be used for  detection of
errors,  but not  for  quantum  error correction.   Such  a code  with
stabilizer structure,  which we call an  ``ambiguous stabilizer code''
(ASC), can nevertheless be useful  for the characterization of quantum
dynamics (CQD). The use  of ASCs can help lower the  size of CQD probe
states used, but at the cost of increased number of operations.
\end{abstract}
\maketitle

\section{Introduction}

In  quantum information  processing, a  quantum error  correcting code
(QECC) is a subspace $\mathcal{C}$, carefully selected to protect from
certain  noise,  any  initial  state  $|\Psi\rangle  \in  \mathcal{C}$
\cite{KL97}.   Let   $\{|j_L\rangle\}$  be   a  $n$-qubit   basis  for
$\mathcal{C}$, encoding  $k$-qubit states  $|j\rangle$, with $0  \le j
\le 2^{k}-1$. Such a  code is a $[[n,k]]$ QECC, where  $k$ is the code
rate.  In this work, we will represent the noise using the error basis
given by elements $E_k$ of the Pauli group $\mathcal{P}_n$, the set of
all possible tensor products of  $n$ Pauli operators, with and without
factors $\pm1, \pm  i$.  Thus $E^\dag_k = E_k$  and $(E_k)^2=I_n$, the
identity operator over $n$ qubits.

\subsection{Stabilizer formalism}

A stabilizer description of error correction is connected to classical
error correcting codes over GF(4) \cite{Got09}.  Where applicable, the
stabilizer  formalism  is   advantageous  in  focussing  attention  on
measurement  operators, which  can  be compact,  rather  than on  code
words, which  can be  large.  A state  $|\psi_L\rangle$ is said  to be
stabilized by  an operator $S$ if  $S|\psi_L\rangle = |\psi_L\rangle$.
Let $\mathcal{G}$ be a subset of $n-k$ independent, commuting elements
from  $\mathcal{P}_n$.   A  $[[n,k]]$  QECC is  the  $2^k$-dimensional
simultaneous   $+1$-eigenspace  $\mathcal{C}$   of  the   elements  of
$\mathcal{G}$.   A   basis  for  $\mathcal{C}$  are   the  code  words
$|j_L\rangle$.    The  set   of  $2^{n-k}$   operators   generated  by
$\mathcal{G}$   constitute    the   stabilizer   $\mathcal{S}$.    The
centralizer  of   $\mathcal{S}$  is  the   set  of  all   elements  of
$\mathcal{P}_n$ that commute with each member of $\mathcal{S}$:
\begin{equation}
\mathcal{Z} = \{P \in \mathcal{P}_n~|~ \forall S \in \mathcal{S},
  [P,S] = 0\},
\end{equation}
while the  normalizer of $\mathcal{S}$ is  the set of  all elements of
$\mathcal{P}_n$ that conjugate the stabilizer to itself:
\begin{equation}
\mathcal{N} = \{P \in \mathcal{P}_n~|~ P\mathcal{S}P^\dag = \mathcal{S}\}.
\end{equation}
We note that $\mathcal{S}  \subseteq \mathcal{N}$ because the elements
of  $\mathcal{S}$  are   unitary  and  mutually  commute.   Similarly,
$\mathcal{Z}   \subseteq   \mathcal{N}$   because  elements   of   the
$\mathcal{Z}$   are  unitary   and  commute   with  all   elements  of
$\mathcal{S}$.   To see that  $\mathcal{N} \subseteq  \mathcal{Z}$, we
note that if  $N \in \mathcal{N}$ then given  any $S \in \mathcal{S}$,
$NSN^\dag = S^\prime$,  for some $S^\prime \in \mathcal{S}$,  or $NS =
S^\prime N$. For  Pauli operators, $NS = \pm  SN$, meaning $S^\prime =
\pm S$.   But if $S^\prime  = -S$, then  $NSN^\dag = -S$,  which would
require that both $S$ and $-S$ are in $\mathcal{S}$.  However if $S\in
\mathcal{S}$,  then  $-S$  is  not  in the  stabilizer,  so  the  only
possibility is $S^\prime = S$, and we obtain that for each $N$ and any
$S$,  $[N,S]=0$, i.e., $\mathcal{N}  \subseteq \mathcal{Z}$.   It thus
follows   that   here    $\mathcal{Z}   =   \mathcal{N}$.    We   have
$SN|j_L\rangle=NS|j_L\rangle=N|j_L\rangle$,   which  shows   that  the
action of $N$  maps code words to code words, and  thus has the action
of a logical Pauli operation on code words.

A set  of operators  $E_j \in \mathcal{P}_n$  constitutes a  basis for
correctable  errors if one of the following conditions hold:
\begin{subequations}
\begin{eqnarray}
E_j E_k &\in& \mathcal{S} \label{eq:deg} \\
\exists G &\in& \mathcal{G}: [E_j E_k,G] \ne 0. \label{eq:nodeg}
\end{eqnarray}
\label{eq:stabqecc}
\end{subequations}
The  case  (\ref{eq:deg})  corresponds to  \textit{degeneracy}.   Here
$\langle\psi_L|E_jE_k|\psi_L\rangle=    \langle\psi_L|\psi_L\rangle=1$,
meaning that both  errors produce the same effect,  and the code space
is  indifferent  as to  which  of  them  happened. Thus  either  error
operator  can be  applied as  a recovery  operation when  one  of them
occurs.   The case  (\ref{eq:nodeg})  corresponds to  $E_j E_k  \notin
\mathcal{N}$.   In that  case,  $\exists G\in\mathcal{G}  : E_jE_kG  =
-GE_jE_k$, which ensures that  $G$ anti-commutes with precisely one of
the  operators  $E_j$  and  $E_k$.   Thus  the  noisy  logical  states
$E_j|\psi_L\rangle$   and  $E_k|\psi_L\rangle$  will   yield  distinct
eigenvalues  (one being  ${+1}$  and  the other  ${-1}$)  when $G$  is
measured.  The  set of $n-k$ eigenvalues $\pm1$  obtained by measuring
the generators  $G$ forms the error syndrome.   The consolidated error
correcting  condition   (\ref{eq:stabqecc})  can  be   stated  as  the
requirement 
\begin{equation}
E_jE_k \notin \mathcal{N}-\mathcal{S},
\label{eq:qecc0}
\end{equation}
for every pair of error basis elements, with $j \ne k$.

\subsection{Noise characterization and stabilizer codes}

Characterization of the quantum dynamics  (CQD) of a quantum system is
vital  for practical applications  in quantum  information processing,
communication   and   computation,   on   account   of   environmental
decoherence. If $\sigma$ represents the quantum state of the system at
time $t=0$, then it evolves under the action of the noise to
\begin{eqnarray}
\mathcal{E}(\sigma) =
\sum_{m,n} \chi_{m,n}E_m \sigma E_n^\dag,
\label{eq:chi}
\end{eqnarray}
where  $\chi$ is  the  \textit{process matrix},  a Hermitian  operator
satisfying  the  properties  $  \sum_{j,k}  \chi_{j,k}  E^\dag_jE_k  =
\mathbb{I}$, and  $\sum_m \chi_{m,m}=1$ \cite{OSB15}.   The (positive)
elements  $\chi_{j,j}$ are  probabilities for  errors $E_i$  to occur.
The terms $\chi_{j,k}$  ($j\ne k$) refer to the  coherence between two
distinct errors.

Quantum  process tomography (QPT)  denotes a  CQD technique  where for
selected  input  states $\sigma_j$,  complete  state tomographic  data
$\lambda_{j,k}   =  \textrm{Tr}(\sigma_j   \mathcal{E}(\sigma_k))$  is
obtained.   The  process  matrix  is  derived  by  inversion  of  this
experimental  data.   There have  been  several  QPT techniques,  like
standard  quantum  process  tomography  (SQPT)  \cite{NC00,DAr00}  and
ancilla-assisted process  tomography (AAPT) \cite{ABJ+03}.   Other CQD
methods,   which    bypass   state   tomography,    include   ``direct
characterization  of quantum  dynamics'' (DCQD)  \cite{ML06,*ML07} and
``quantum  error  correction   based  characterization  of  dynamics''
(QECCD),  introduced  by  the  present  authors  \cite{OSB15}.   Other
developments include an efficient method for estimating diagonal terms
of the  process matrix using  twirling \cite{ESM+07}, which  is useful
for determining  QECCs \cite{SMK+08}.  Other related  works on channel
estimation include  Ref. \cite{BPP08}, a  technique like that  in Ref.
\cite{SMK+08}  extended to cover  off-diagonal $\chi_{jk}$  terms, and
Refs. \cite{FSK+14,*CFC+14,*Fuj14}.

QECCD brings  about a twist  to the theme  of using QECCs in  that the
codes are  used not just for  protecting quantum states,  but also for
CQD.   This permits  one to  implement CQD  concurrently  with quantum
computation, making  QECCD work ``online''. The intuition  is that the
statistics of errors detected  during the error correction process are
used to characterize the noise.   QECCD makes use of the properties of
a class of stabilizer codes in which the allowed Pauli error operators
form a group.

\subsection{Code ambiguity}

A natural  extension of  the concept  behind QECCD  would be  to adapt
stabilizer techniques  purely or  primarily for  CQD, rather  than for
quantum error  correction. Freed  from correction  duty, codes  are no
longer bound  to obey (\ref{eq:stabqecc}).   This can be  exploited to
design  codes  with  code   lengths  smaller  than  permissible  under
(\ref{eq:stabqecc}),  thereby making  code words  easier to  implement
experimentally.

However the  price to  pay for  violating (\ref{eq:stabqecc})  is that
some errors  will be indistinguishable, making  stabilizer measurement
outcomes  ambiguous.   The  new  class of  stabilizer  codes  that  we
introduce  here  are  therefore  called  \textit{ambiguous  stabilizer
  codes}  (ASCs).  In  an  ASC,  the final  state  after recovery  may
contain a residual logical Pauli operation with respect to the initial
logical state.  An  ASC generalizes the concept of  a degenerate code,
which is  the special case  where the only residual  logical operation
after recovery is the trivial one.

Thus,  the   purpose  of  invoking  ambiguity--indeed   the  principal
motivation  behind  the construction  of  ASC's--  is to  exploit  the
stabilizer formalism and structure for  the construction of codes that
are better  suited for  error characterization  rather than  for error
correction. 

Code   ambiguity  entails,   as  detailed   below,  that   more  state
preparations  involving  other  ASCs  are  required  to  unambiguously
determine the  process matrix. Later we  will find that the  number of
ASCs  required   for  full   characterization  scales   linearly  with
ambiguity, in  a way made  precise later.   Thus there is  a trade-off
between  spatial  resources  (length   of  code  words)  and  temporal
resources (number of operations).  We call this ambiguous extension of
QECCD as  ``quantum ASC-based characterization of  dynamics'' (QASCD).
From an experimental viewpoint, the  above trade-off means QASCD helps
simplify quantum state preparation at  the cost of increased number of
trials and classical post-processing.

QASCD is,  unlike traditional methods  of process tomography  but like
the techniques presented in Refs. \cite{ML06,OSB15}, a \textit{direct}
method in that it does not  require the full state tomography of probe
states  used for CQD.   At the  same time,  QASCD may  require smaller
codes  than these  direct techniques.  In particular,  to characterize
$m$-qubit  noise, the  above direct  techniques require  probes  to be
$2m$-qubit states or larger.  By contrast, with ASCs one can beat this
bound.   For  example,  to  characterize  2-qubit noise,  one  can  in
principle use (a family of) 3-qubit ASCs.

The  remaining   article  is  structured  as   follows.   After  first
developing a  theory of  ASCs in Section  \ref{sec:ambi}, we  study in
Section  \ref{sec:AG} their  specific  group  theoretic properties  as
would  be useful  for CQD.   In  Section \ref{sec:Xr},  we detail  the
protocol  that would  be  used  for CQD  by  employing  (a family  of)
ASCs.  The  resources, in  terms  of  number  of ASCs  and  operations
required for CQD, are discussed in Section \ref{sec:res}.  A trade-off
between the space  resources (length of codes used)  vs time resources
(required configurations)  is discussed  here. After  illustrating our
new   method  as   applied  to   a  toy   2-qubit  noise   in  Section
\ref{sec:example},   we  finally   present   conclusions  in   Section
\ref{sec:conclu}. 

\section{Ambiguous stabilizer codes \label{sec:ambi}}

\subsection{Definition and basic features}

A  $2^k$-dimensional  subspace  $\mathcal{C}^\prime$  of  $n$  qubits,
together with an allowed set $\mathbb{E}$ of Pauli error operators, is
\textit{ambiguous} when two or more errors cannot be distinguished via
syndrome  measurements on  the logical  state.  The  indistinguishable
errors  may require  different  recovery  operations.  Thus  ambiguity
generalizes the concept  of degeneracy, and in  general prevents error
correction.

An \textit{ambiguous set}  $A^{(p)}$ is a collection  of allowed Pauli
errors   that  cannot   be  mutually   distinguished  by   a  syndrome
measurement.   Let  $A^{(p)}  \equiv \{E_1^{(p)},  E_2^{(p)},  \cdots,
E_{\upsilon(p)}^{(p)}\}$, where $E_j^{(p)}$ is  the $j$th error in the
$p$th ambiguous set, and there are $\upsilon(p)$ elements in the $p$th
ambiguous set. Then any two  errors $E_j^{(p)}$ and $E_k^{(p)}$, where
$j\ne k$,  produce the same  error syndrome.  Thus, Pauli  errors that
are elements of the same ambiguous set are mutually indistinguishable.
However, they  may not be  degenerate, and thus may  require different
recovery  operations, making  an ambiguous  code unsuitable  for error
correction.

Ambiguity of the code can  be represented by partitioning $\mathbb{E}$
into  ambiguous sets.   The collection  of all  ambiguous sets  is the
ambiguous   class   $\mathcal{A}   =   \{A^{(1)},   A^{(2)},   \cdots,
A^{(\sigma)}\}$,    with    $\bigcup_p    A^{(p)}=\mathbb{E}$.     The
\textit{order of  ambiguity} of  the code is  $\sigma$, the  number of
ambiguous  sets  in  the   ambiguous  class.   The  \textit{degree  of
  ambiguity}, denoted $\gamma$, is the  largest number of Pauli errors
in $\mathbb{E}$ that can be mutually ambiguous, i.e.,
\begin{equation}
\gamma\equiv\max_p |A^{(p)}|.
\label{eq:gamma}
\end{equation}
A  conventional  (unambiguous)  stabilizer code  is  characterized  by
$\gamma=1$, whereas  for an  ASC, $\gamma>1$.   Any set  of up  to $s$
known  errors drawn  from  distinct ambiguous  sets  $A^{(p)}$ can  be
detected, and if the errors are known, they can be corrected.

Within an ambiguous set $A^{(p)}$, the error elements produce the same
error syndrome.  This means  that the action  of two  ambiguous errors
$E_n^{(p)}$ and $E_m^{(p)}$ must be related by
\begin{equation}
E_n^{(p)}|j_L\rangle = NE^{(p)}_m|j_L\rangle,
\label{eq:ambi}
\end{equation}
where $N$ is a normalizer  element. Note that $NE^{(p)}_m|j_L\rangle =
\pm E^{(p)}_mN|j_L\rangle$.  Thus,
\begin{eqnarray}
\sum_{j=0}^{2^k-1}    E^{(p)}_n|j_L\rangle\langle   j_L|E_n^{(p)}    &=&
\sum_{j=0}^{2^k-1} E^{(p)}_mN|j_L\rangle\langle j_L|NE_m^{(p)} \nonumber\\
&=&\sum_{j^\prime=0}^{2^k-1}     E^{(p)}_m|j^\prime_L\rangle \langle j^\prime_L|E_m^{(p)} \nonumber\\
&\equiv& \Pi^{(p)},
\label{eq:norm1}
\end{eqnarray}
where  $j^\prime$ is  simply a  re-ordering of  $j$.  In  other words,
every element $E^{(p)}_m$ generates  the same erroneous subspace, with
projector  $\Pi^{(p)}$.   However,   individual  code  words  are  not
necessarily  mapped  to the  same  erroneous  code  word, in  view  of
(\ref{eq:ambi}).   Further, from  Eq.  (\ref{eq:ambi}),  we have  $N =
E^{(p)}_nE^{(p)}_m$.     If     $E_m^{(p)}|j_L\rangle    =    N^\prime
E^{(p)}_n|j_L\rangle$,  then $N^\prime  =  E^{(p)}_mE^{(p)}_n$.  Thus,
$N^\dag = N^\prime$.

Note that if $[E^{(p)}_n,E^{(p)}_m] = 0$, then $N^\dag = N$ (Hermicity
condition) and  thus $N=N^\prime$.  Conversely, if  $N=N^\prime$, then
$E^{(p)}_nE^{(p)}_m      =      E^{(p)}_mE^{(p)}_n$,     and      thus
$[E^{(p)}_n,E^{(p)}_m]=0$.   If  $\{E^{(p)}_n,E^{(p)}_m\} =  0$,  then
$N^\dag = -N$ (anti-Hermiticity)  and thus $N=-N^\prime$.  Conversely,
if $N=-N^\prime$, then $E^{(p)}_nE^{(p)}_m = -E^{(p)}_mE^{(p)}_n$, and
thus $\{E^{(p)}_n,E^{(p)}_m\}=0$.

In contrast to the case  with subspaces generated by ambiguous errors,
projectors to distinct unambiguous erroneous subspaces are orthogonal:
\begin{equation}
 \Pi^{(p)}\Pi^{(q)}   = 0,
\label{eq:completeness}
\end{equation}
if $p \ne q$. Thus two  or more errors belonging to distinct ambiguous
sets can always be disambiguated.

\subsection{Ambiguously detectable errors}

Ambiguous errors  $E^{(p)}_m$ and $E^{(p)}_n$  that are linked  in Eq.
(\ref{eq:ambi})  with  $N=I_L$,  where  $I_L$  is  the  logical  Pauli
identity  operator, require  the same  recovery operation.   Ambiguous
errors related  by non-trivial  logical Pauli operations  will require
distinct recovery operations.  Thus, an  ambiguous code cannot be used
for quantum error correction.

For ASCs, the error correcting condition (\ref{eq:stabqecc}) becomes:
\begin{subequations}
\begin{eqnarray}
p \ne q ~\Rightarrow~ E_m^{(p)} E_n^{(q)} &\notin& 
\mathcal{N} \label{eq:a-deg}, \\
p = q ~\Rightarrow~ E_m^{(p)} E_n^{(q)} &\in& 
\mathcal{N}.
 \label{eq:a-nodeg}
\end{eqnarray}
\label{eq:a-stabqecc}
\end{subequations}
Eq.   (\ref{eq:a-deg}) implies  that quantum  error correction  can be
implemented for  any collection of \textit{known}  errors which belong
to distinct  ambiguous sets.  Eq. (\ref{eq:a-nodeg})  implies that any
pair of  errors belonging to the  same ambiguous set will  produce the
same  syndrome,  and thus  be  indistinguishable.   In particular,  if
$E_m^{(p)} E_n^{(p)}  \in \mathcal{S}$,  then $\langle\psi_L|E_m^{(p)}
E_n^{(p)}|\psi_L\rangle  = \langle\psi_L|\psi_L\rangle$,  meaning that
the two errors are mutually  degenerate, and the ambiguity is harmless
in the sense that the recovery operation for any one of them works for
the  other, too.   On  the  other hand,  if  $E_m^{(p)} E_n^{(p)}  \in
\mathcal{N}-\mathcal{S}$, then  the erroneous code words  they produce
are   related   by   non-trivial    logical   Pauli   operations   Eq.
(\ref{eq:ambi}), and the  error correcting conditions (\ref{eq:qecc0})
are violated. If one implements a recovery operation favoring a single
error in each ambiguous set, this will in general produce a mixture of
states within  the code space $\mathcal{C}^\prime$,  which are logical
Pauli rotated versions of each other.

In $\mathcal{A}$, each ambiguous set $A^{(p)}$ corresponds to the same
error syndrome, so that order $\sigma \le 2^{n-k}$.  By definition, the set
$A^{(0)}$   will  contain   the  element   $I$  and,   by  virtue   of
Eq. (\ref{eq:a-nodeg}), only elements of the normalizer $\mathcal{N}$.
The  remaining  sets  $A^{(1)},A^{(2)},  \cdots$  will  contain  Pauli
operators  not  present in  $\mathcal{N}$,  since  they will  fail  to
commute with at least one  stabilizer generator.  

For unambiguous (and non-degenerate) recovery using a linear QECC, the
dimension of the code space and the volume $|\mathbb{E}|$ must satisfy
the quantum Hamming bound, $2^k |\mathbb{E}| \le 2^{n}$, or
\begin{equation} 
\log(\mathbb{E}) \le n-k.
\label{eq:QHB}
\end{equation}
An ambiguous code may  violate (\ref{eq:QHB}), though not necessarily.
A QECC  that saturates Eq. (\ref{eq:QHB})  is called \textit{perfect}.
The 5-qubit code of Ref.  \cite{LMP+96} is such an example.

\subsection{Constructing ASC's}

The simplest way to produce  an ASC is by \textit{error overloading} a
stabilizer  code.    This  involves  allowing   additional  errors  in
violation   of  condition   (\ref{eq:stabqecc}),  such   that  instead
condition  (\ref{eq:a-stabqecc}) holds.   Ambiguity produced  by error
overloading a perfect  code will result in a  violation of the quantum
Hamming  bound   (\ref{eq:QHB}),  while  for  an   imperfect  QECC,  a
sufficiently large  amount of error  overloading would be  required to
violate  Ineq. (\ref{eq:QHB}).   For example,  consider  the (perfect)
5-qubit code of Ref. \cite{LMP+96}
\begin{eqnarray}   |0_L\rangle_5    &=&    \frac{1}{2\sqrt{2}}
\left(-|00000\rangle  + |0111\rangle  - |10011\rangle  + |11100\rangle
\right.    \nonumber\\   &&\left.+|00110\rangle   +  |01001\rangle   +
|10101\rangle  + |11010\rangle\right)\nonumber\\  |1_L\rangle_5 &  = &
\frac{1}{2\sqrt{2}}\left(-|11111\rangle                               +
|10000\rangle+|01100\rangle-|00011\rangle
\right.\nonumber\\    &&\left.+|11001\rangle    +   |10110\rangle    -
|01010\rangle-|00101\rangle\right),
\label{eq:5qecc}
\end{eqnarray} 
which corrects an arbitrary single-qubit error on any qubit.  The code
space is stabilized by  generators $IXXYY, IYYXX, XIYZY, YXYIZ$.  They
can  each  take  values  $\pm1$,  thereby  determining  16  syndromes,
corresponding to  the 16 allowed  errors $\mathbb{E} \equiv  \{I, X_i,
Y_i, Z_i\}$ where $i=1,\cdots,5$. By allowing any more errors into the
error  set  $\mathbb{E}$, we  introduce  ambiguity,  and also  violate
(\ref{eq:QHB}).   In  Table  \ref{tab:5aqecc},  we present  a  partial
listing of the ambiguous class $\mathcal{A}$ for this code. In all, it
has $1 + {5 \choose 1}\cdot 3 + {5 \choose 2}\cdot3^2 = 106$ arbitrary
1-qubit and 2-qubit errors, of which 49 are displayed.  The errors are
partitioned  into their  respective  ambiguous sets,  labelled by  the
corresponding error syndrome.  Set  $A^{(0)}$ has only 1 element, $I$,
since  all  other elements  of  $\mathcal{N}$  have  a Hamming  weight
greater than 2.

\begin{center}
\begin{table}
\begin{tabular}{|l|l|l|l|l|l|l|l|l|}\hline
++++    & +++ $-$    & ++$-$ +  & ++$--$  \\\hline
$I$     & $X_1$      & $Y_1$    & $Z_1$ \\
        & $Y_2Y_3$   & $Z_2Z_3$ & $X_2X_3$ \\
        & $X_3Y_4$   & $Y_3X_4$ & $Z_3Z_4$ \\\hline
 +$-$++ &+$-$+$-$   &+$--$+    &+$---$ \\ \hline 
  $X_2$   &$Y_5$   &$Y_4$   &$X_3$ \\
 $Z_1X_3$&$X_1X_2$&$Y_1X_2$&$Z_1X_2$ \\
 $Y_3Z_4$&$Z_2Y_3$&$Y_2Z_3$&$Z_2X_4$ \\ \hline 
$-$+++   & $-$++ $-$   & $-$+$-$ + & $-$+$--$ \\\hline
$Y_3$    & $Y_2$       & $X_4$     & $X_5$    \\
$X_1Y_2$ & $X_1Y_3$    & $Z_1Y_2$  & $Y_1Y_2$\\
$X_2Z_4$ & $Z_3Y_4$    & $Z_2X_3$  & $X_2Z_3$ \\\hline
 $-$$-$++ &$-$$-$+$-$&$-$$--$+ &$-$$---$ \\ \hline 
  $Z_4$    &$Z_2$     &$Z_5$    &$Z_3$ \\
 $X_1Z_2$ &$Y_1Z_3$  &$Z_1Z_2$ &$Y_1Z_2$ \\
 $X_2Y_3$ &$X_3X_4$  &$X_1Z_3$ &$Y_2Y_4$ \\ \hline
\end{tabular} 
\caption{Ambiguous  class (partial  listing) for  the ASC  obtained by
  error-overloading  the  code  (\ref{eq:5qecc}), to  allow  arbitrary
  errors on any  two qubits.  Each error syndrome  labels an ambiguous
  set.  The  first error row  in each column corresponds  to arbitrary
  single-qubit errors allowed in the  original QECC.  Inclusion of the
  two-qubit errors  (second and third rows  of the table) to  the list
  turns the QECC into  an ASC.  In all, there are  106 elements in the
  ambiguous   class,   with   $|A^{(0)}|=1$  and   $|A^{(p)}|=7$   for
  $p=1,2,\cdots,15$.  Thus the degree of  ambiguity is 7. For example,
  the full ambiguous  set, corresponding to the  syndrome ${+++-}$ has
  four  more  elements  $E^{(1)}_3  \equiv  X_4X_5,  E^{(1)}_4  \equiv
  Z_3Z_5,  E^{(1)}_5 \equiv  X_2Y_5$, and  $E^{(1)}_6 \equiv  Z_2Z_4$.
  The normalizers between $E^{(1)}_0 \equiv X_1$ and other elements in
  the  set  are $XYYII  \rightarrow  Z_L$,  $XIXYI \rightarrow  -Y_L$,
  $XIIXX   \rightarrow   Z_L$,   $XIZIZ  \rightarrow   -X_L$,   $XXIIY
  \rightarrow -Y_L$ and $XZIZI \rightarrow  -X_L$.  Any set of sixteen
  elements,  with  one drawn  from  each  ambiguous set  will  satisfy
  condition Eq.  (\ref{eq:a-deg}),  while any pair of  errors within a
  column    satisfy    Eq.     (\ref{eq:a-nodeg})   and    thus    are
  ambiguous. Further note that the  product of ambiguous errors linked
  by the same logical Pauli  are mutually degenerate (e.g., $E^{(1)}_4
  E^{(1)}_6  \in  \mathcal{S}$),  and  are  correctable  by  the  same
  recovery operation,  while those  linked by different  logical Pauli
  operators    are    not     (e.g.,    $E^{(1)}_4    E^{(1)}_5    \in
  \mathcal{N}-\mathcal{S}$).}
\label{tab:5aqecc}
\end{table}
\end{center}

Another  way  to   create  an  ASC  from  a   stabilizer  code  is  by
\textit{syndrome  coarse-graining}:  dropping  one  or  more  syndrome
measurements. For  example consider  \textit{not} to measure  the last
stabilizer  of  the  code  (\ref{eq:5qecc}).   From  the  first  entry
corresponding to syndromes  (the \textit{un}-error-overloaded case) of
the  Table \ref{tab:5aqecc}  it  can be  seen that  $|\mathcal{A}|=8$,
$A^{(0)}=\{I,X_1\}$     corresponding     to     syndrome     $(+++)$,
$A^{(1)}=\{Y_1,Z_1\}$  corresponding to syndrome  $(++-)$, and  so on.
The  order of  ambiguity  is halved  and  the degree  of ambiguity  is
doubled.

A final  method to obtain an  ASC begins by constructing  a stabilizer
code that corrects arbitrary errors  on $m$ known coordinates.  An ASC
may then  be obtained  by allowing  noise to  act on  $m^\prime$ known
coordinates,  where  $m^\prime>m$.   A detailed  description  of  this
method and its application to the characterization of quantum dynamics
\cite{OSB15} are considered below.

\section{Ambiguous group \label{sec:AG}}

 An  arbitrary  error on  $l$  qubits can  be  expressed  as a  linear
 combination of $4^l$ Pauli  operators.  Suppose these $l$-qubits form
 a subsystem  of $n$ qubits  prepared in a $[[n,k]]$  stabilizer code.
 Setting $|\mathbb{E}| := 4^l$ in Ineq. (\ref{eq:QHB}) we find:
\begin{equation}
l \le \lfloor \frac{n-k}{2}\rfloor
\label{eq:newQHB}
\end{equation}
This means that  a 5-qubit code can correct all  possible errors on at
most 2  fixed coordinates. An example  of a perfect code  of this kind
will be presented  later.  We thus obtain a  $[[n,k]]$ ASC by allowing
$m$     noisy     coordinates,     where    $m>l$     in
Ineq. (\ref{eq:newQHB}).  The order $\sigma$  of the code is  just the
number of syndromes, $2^{n-k}$,  while the degree of ambiguity $\gamma
=  4^{m}/2^{n-k}   =  2^{2m-n+k}$.

Suppose we are given a $[[n,k]]$  ASC with errors allowed on $m$ known
coordinates.   It  is  worth  noting  here  that  the  set  of  errors
(including   the  factors   $\pm1,\pm   i$)  forms   a  group,   i.e.,
$\mathbb{E}=\mathcal{P}_m$.  Furthermore, the subset $\mathfrak{B}$ of
$\mathcal{P}_m$ that is ambiguous with $I_m$ (the trivial error on the
$m$  qubits) constitutes  a  group, the  \textit{ambiguous group},  as
shown below.
\begin{thm}
Given  a $[[n,k]]$  ASC  with  $\mathbb{E}=\mathcal{P}_m$, the  subset
$\mathfrak{B}$  of  allowed errors  that  correspond  to the  no-error
syndrome forms a normal subgroup.
\label{thm:group}
\end{thm}
\textbf{Proof.}  Note that if  $B_j, B_k \in \mathfrak{B}$, then $B_j$
and  $B_k$  both  commute  with  all stabilizers,  by  virtue  of  Eq.
(\ref{eq:a-nodeg}).     (Note   that    this   doesn't    imply   that
$[B_j,B_k]=0$. Thus the  subgroup is not Abelian.) For  any element $G
\in \mathcal{G}$, then $[B_jB_k, G]  = B_jB_kG - GB_jB_k = 0$, meaning
that $B_jB_k  \in \mathfrak{B}$.  This guarantees closure  of the set.
By definition, $I_m$  is an element of this set,  and a Pauli operator
is its own inverse.  Thus all required group properties are satisfied.
Normalcy of the  subgroup (the equality of the  left and right cosets)
is guaranteed  because we implicitly include Pauli  operators with and
without factors $\pm1$ and $\pm i$. \hfill $\blacksquare$
\bigskip

For an ASC obtained in this way, the ambiguous class $\mathcal{A}$ has
a simple structure. It  corresponds to a partition of $\mathcal{P}_m$,
determined by the quotient or factor group
\begin{equation}
\mathcal{Q} \equiv \frac{\mathcal{P}_m}{\mathfrak{B}}.
\label{eq:Q}
\end{equation}
This  means that  any  element  $E$ in  $\mathcal{P}_m$  is either  in
$\mathfrak{B}$ or  can be expressed  as the  product of an  element in
$\mathfrak{B}$ and an element not in $\mathfrak{B}$.

\subsection{Example of a $[[3,1]]$ ASC}

A  $[[3,1]]$ perfect QECC  that unambiguously corrects  errors on
the first qubit is \cite{OSB15}:
\begin{eqnarray}
|0_L\rangle_3 &=& \frac{1}{2}(|001\rangle  + |010\rangle + |100\rangle +
|111\rangle)  \nonumber \\  
|1_L\rangle_3 &=&  \frac{1}{2}(|110\rangle -
|101\rangle + |011\rangle - |000\rangle),
\label{eq:3qec}
\end{eqnarray}
whose stabilizer generators are given by the set $\mathcal{G}_3 \equiv
\{XIX,  YYZ\}$.  The  stabilizer is  thus  the set  of four  elements,
$\mathcal{S}_3 = i^4 \times 2^\mathcal{G}  \equiv i^4 \times \{I, XIX,
YYZ, ZYY\}$,  where the  pre-factor indicates possible  factors $\pm1,
\pm i$.  The normalizer $\mathcal{N}_3$ is  the set of all elements of
$\mathcal{P}_3$  that commute  with the  elements of  $\mathcal{S}_3$.
(We note  that a  Pauli operator  $P$ commutes  with every  element of
$\mathcal{S}_3$   iff    $P$   commutes   with   every    element   of
$\mathcal{G}_3$.)

\begin{table}[h]
$$
\begin{array}{c|c|c|c}\hline
I_L & -X_L & Y_L & Z_L \\
\hline
III & XZI & IYI & XXI \\
XIX & IZX & XYX & IXX \\
YYZ & YXY & YIZ & YZY \\
ZYY & ZXZ & ZIY & ZZZ \\
\hline
\end{array}
$$
\caption{Normalizer for the  $[[3,1]]$ stabilizer code (\ref{eq:3qec})
  and the  logical Pauli operations  they map to. All  elements commute
  with  the  elements  of  $\mathcal{S}_3$,  while  their  commutation
  properties amongst themselves reflect the logical operation they map
  to. Thus, an element in the  column $Y_L$ commutes with all elements
  in  the  same  column  and  those in  the  column  $I_L$,  but  will
  anti-commute with every element in  the columns -$X_L$ and $Z_L$. On
  the  other  hand,  the  elements  in the  column  $I_L$,  which  are
  precisely those of $\mathcal{S}_3$, commute with every other element
  in the normalizer.}
\label{tab:nor3}
\end{table}

For code  (\ref{eq:3qec}), the normalizer $\mathcal{N}_3$  is given in
Table  \ref{tab:nor3}.  Since  there   are  only  four  logical  Pauli
operator, various  normalizer elements map  to the same  logical Pauli
operator by virtue of their  effect on the code words (\ref{eq:3qec}).
The  subset  $\mathcal{S}_3$ (the  first  column)  corresponds to  the
identity  logical  Pauli  operation   $I_L$,  while  the  elements  of
$\mathcal{N}_3-\mathcal{S}_3$ correspond to  non-trivial logical Pauli
operations,   as  tabulated   in  the   remaining  columns   of  Table
\ref{tab:nor3}.

We  create an  ASC  for  the code  (\ref{eq:3qec}) by  allowing
errors,  in addition  to  the  first coordinate,  also  on the  second
coordinate. There are four elements  in Table \ref{tab:nor3} that have
no non-trivial operator on the last  qubit, i.e., they are elements of
$\mathcal{P}_2  \otimes  \mathbb{I}_3$,  where $\mathbb{I}_3$  is  the
identity  operator on  the third  qubit.  They  are $\{III,  XZI, IYI,
XXI\}$, which  constitute the  ambiguous group  $\mathfrak{B}_3$.  The
partitioning   of   $\mathcal{A}$   for   ASC   (\ref{eq:3qec})   with
$\mathbb{E}=\mathcal{P}_2$ can be represented by the quotient group:
\begin{equation}
\mathcal{Q}_3 \equiv \frac{\mathcal{P}_2}{\mathfrak{B}_3}.
\label{eq:qg3}
\end{equation}
This is depicted  in Table \ref{tab:3}, where the first  column is the
ambiguous  subgroup $\mathfrak{B}_3$,  and the  other columns  are its
cosets and the other ambiguous errors.

\begin{table}
\begin{tabular}{|l|l|l|l|l|l|}\hline
++          & + $-$       & $-$ +       & $--$        & Normalizer \\\hline
$I$         & $X_1$       & $Y_1$       & $Z_1$       & $I_L$ \\
$Y_2$       & $X_1Y_2$    & $Y_1Y_2$    & $Z_1Y_2$    & $Y_L$ \\
$X_1X_2$    & $X_2$       & $Z_1X_2$    & $Y_1X_2$    & $Z_L$ \\
$X_1Z_2$    & $Z_2$       & $Z_1Z_2$    & $Y_1Z_2$    & $-X_L$ \\\hline
%$X_2X_3$    & $X_1X_2X_3$ & $Y_1X_2X_3$ & $X_1X_2X_3$ &\\
%$Z_2X_3$    & $X_1Z_2X_3$ & $Y_1Z_2X_3$ & $_1Z_2X_3$  &\\
%$X_1X_3$    & $X_3$       & $Z_1X_3$    & $Y_1X_3$    &\\
%$X_1Y_2X_3$ & $Y_2X_3$    & $Z_1Y_2X_3$ & $Y_1Y_2X_3$ &\\
%$Y_1Z_3$    & $Z_1Z_3$    & $Z_3$       & $X_1Z_3$    &\\
%$Y_1X_2Y_3$ & $Z_1X_2Y_3$ & $X_2Y_3$    & $X_1X_2Y_3$ &\\
%$Y_1Y_2Z_3$ & $Z_1Y_2Z_3$ & $Y_2Z_3$    & $X_1Y_2Z_3$ &\\
%$Y_1Z_2Y_3$ & $Z_1Z_2Y_3$ & $Z_2Y_3$    & $_1Z_2Y_3$  &\\
%$Z_1Y_3$    & $Y_1Y_3$    & $X_1Y_3$    & $Y_3$       &\\
%$Z_1X_2Z_3$ & $Y_1X_2Z_3$ & $X_1X_2Z_3$ & $X_2Z_3$    &\\
%$Z_1Y_2Y_3$ & $Y_1Y_2Y_3$ & $X_1Y_2Y_3$ & $Y_2Y_3$    &\\
%$Z_1Z_2Z_3$ & $Y_1Z_2Z_3$ & $X_1Z_2Z_3$ & $Z_2Z_3$    &    \\\hline
\end{tabular} 
\caption{Ambiguous  class $\mathcal{A}_3$  for errors  on the  first 2
  qubits of 3-qubit code (\ref{eq:3qec}), depicting the quotient group
  (\ref{eq:qg3}).    The  first   column   is   the  ambiguous   group
  $\mathfrak{B}_3$, drawn  from Table  \ref{tab:nor3}, subject  to the
  requirement that the  operator on the third qubit  is identity.  The
  remaining   three  columns   are   its  cosets   $X_1\mathfrak{B}_3,
  Y_1\mathfrak{B}_3$   and    $Z_1\mathfrak{B}_3$,   which   represent
  ambiguous sets.  The  last column lists the  normalizer element with
  respect  to   first  element  in   the  column,  in  the   sense  of
  Eq. (\ref{eq:ambi}).  For example,  the normalizer element that maps
  error $Z_1Z_2$  to error  $Y_1$ or vice  versa is  $N_{Z_1Z_2,Y_1} =
  N_{Y_1,Z_1Z_2}  = -X_L$,  while  the normalizer  element which  maps
  error $Y_1Y_2$ to error $Z_1X_2$ is $N_{Z_1X_2,Y_1Y_2} = Z_LY_L = iX_L$,
  while  $N_{Y_1Y_2,Z_1X_2} = Y_LZ_L = -iX_L$.}
\label{tab:3}
\end{table}

\section{Application to noise characterization \label{sec:Xr}}

While  code  ambiguity  makes  ASCs   not  useful  for  quantum  error
correction, they  can be used  for experimentally studying  noise.  In
this  Section, we  elaborate on  the intuition  presented earlier,  of
extending QECCD by replacing the use  of stabilizer codes with that of
ASCs.

\subsection{Ambiguity and QEC channel-state isomorphism \label{sec:char}}

The basis of QECCD is  the quantum error correction (QEC) isomorphism,
qualitatively   similar   to    the   Choi-Jamiokowski   channel-state
isomorphism,  which associates  a correctable  noise channel  with the
unique erroneous logical state corresponding  to a given input logical
state. This  clearly is  necessary if  complete information  about the
channel  is to  be extracted  via  measurements.  In  the presence  of
ambiguity, for any initial logical state, it can be shown that one can
always construct two or more noise channels such that they produce the
same erroneous logical state. Thus QEC isomorphism no longer holds.

In  QECCD,  the  QEC  isomorphism  is  leveraged  through  some  state
manipulations to  yield full noise  data. The  basic idea is  that the
syndrome obtained from  the stabilizer measurement is  used to correct
the  noisy state,  while the  experimental probabilities  of syndromes
will  characterize the  noisy quantum  channel. While  direct syndrome
measurements  yield the  diagonal  terms of  the  process matrix,  for
off-diagonal terms  preprocessing via suitable unitaries  is required.
For the purpose of noise characterization, the code qubits are divided
into two parts;  (a) the qubits on which the  elements of $\mathbb{E}$
act non-trivially; (b) the remaining qubits.

The former  qubits constitute  the principal system  \textbf{P}, whose
unknown dynamics is to be determined. The latter qubits constitute the
CQD ancilla \textbf{A}, and are  assumed to be clean, i.e., noiseless.
Suppose the  full system $\textbf{P}+\textbf{A}$  is in the  state \be
|\psi_L\rangle \equiv\sum_{j=0}^{2^k-1} \alpha_j|j_L\rangle, \ee where
$\{|j_L\rangle\}$  denotes a  logical basis  for the  code space  of a
$[[p+q,k]]$ ASC (which encodes $k$  qubits into $n \equiv p+q$ qubits)
such that  allowed errors in  the $p$ known coordinates  of \textbf{P}
can be  ambiguously detected.  

The main  difference QASCD  has with  respect to  QECCD is  that QASCD
employs more than one code to fully characterize the noise. Herebelow,
we present  details of the QASCD  protocol, which has a  quantum part,
which  is   experimental,  and   a  classical  part,   which  concerns
post-processing  data from  experiments.   The  quantum part  involves
using state  preparations and syndrome measurements  of different ASCs
to determine  $\chi_{m,n}$ ambiguously.   The classical  part involves
simultaneous  equations  to  disambiguate the  ambiguous  experimental
probabilities.

\subsection{Direct measurement\label{sec:diag}}

Let $Q$ be an ASC that can detect noise $\mathcal{E}$, with
associated    process     matrix    $\chi$.      Let    $E_{\alpha_j}$
($j=0,1,2,\cdots,\gamma-1$) be the elements of an ambiguous set  in $Q$,
with $E_x$ denoting any one of  these $\alpha_j$'s.  It is convenient to
employ the notation $\left|j^{(\alpha)}_L\right\rangle \equiv E_\alpha|j_L\rangle$.
The  probability  that one of  these  ambiguous errors  occur:
%\begin{widetext}
\begin{eqnarray} 
&&\hspace{-0.75cm}\xi\left(\bigwedge_j \alpha_j\right) =
\textrm{Tr}\left(\mathcal{E}\left(
    |\psi_L\rangle\langle\psi_L|\right)
   \left[\sum_{j=0}^{2^k-1}|j_L^x\rangle\langle
  j_L^x|\right] \right) \nonumber \\
 &=& \sum_{j=0}^{2^k-1}
 \left\langle j_L^{x}\right| \left[\cdots + 
\chi_{\alpha_1,\alpha_1}|\psi_L^{(\alpha_1)}\rangle\langle\psi_L^{(\alpha_1)}| 
+
\chi_{\alpha_1,\alpha_2}|\psi_L^{(\alpha_1)}\rangle\langle \psi_L^{(\alpha_2)}| 
\nonumber \right. \\ && +~  
\chi_{\alpha_2,\alpha_1}|\psi_L^{(\alpha_2)}\rangle\langle \psi_L^{(\alpha_1)}| 
 + \left.
\chi_{\alpha_2,\alpha_2}|\psi_L^{(\alpha_2)}\rangle\langle \psi_L^{(\alpha_2)}| + \cdots
   \right]
    |j_L^{x}\rangle \nonumber \\
&=& \cdots + 
\chi_{\alpha_1,\alpha_1} 
+
\chi_{\alpha_1,\alpha_2}\langle \psi_L^{(\alpha_2)}|\psi_L^{(\alpha_1)}\rangle
+
\chi_{\alpha_2,\alpha_1}\langle \psi_L^{(\alpha_1)}|\psi_L^{(\alpha_2)}\rangle 
\nonumber\\ &&+~  
\chi_{\alpha_2,\alpha_2}  + \cdots
   \nonumber \\
&=&
\cdots + 
\chi_{\alpha_1,\alpha_1} 
+
\chi_{\alpha_1,\alpha_2} \langle\psi_L|N_{1,2} |\psi_L\rangle
 +  
\chi_{\alpha_2,\alpha_1}\langle \psi_L|N_{2,1}|\psi_L\rangle
\nonumber \\ && +~
\chi_{\alpha_2,\alpha_2} + \cdots\nonumber\\
&=&  \sum_j \chi_{\alpha_j,\alpha_j}  
+ 2 \sum_{j\ne k} \textrm{Re}\left(\chi_{\alpha_j,\alpha_k}
\langle N_{j,k}\rangle_{L}\right),
\label{eq:diag}
\end{eqnarray}  
%\end{widetext}
where  $N_{m,n}\equiv E_m  E_n$.  Note that  because  $E_m$ and  $E_n$
produce the same syndrome by virtue of ambiguity, $N_{m,n}$ so defined
will commute with all elements of the stabilizer.

Let $D\equiv 2^p$, the dimension of  {\bf P}.  In an unambiguous code,
the $D^2$  diagonal terms of  $\chi$ would appear as  probabilities of
syndrome  measurements \cite{OSB15}.   Now,  however, any  measurement
outcome probability will contain  contributions from the probabilities
of $\gamma$  ambiguous errors  plus ${\gamma \choose  2}$ off-diagonal
terms  between   these  ambiguous  errors.    Of  the  $4^k$   can  be
disambiguated by  using as many different  initial state preparations,
by exploiting  the fact that  the $\chi$  terms have factors  given by
expectation  values of  different normalizer  elements (logical  Pauli
operations).   However,  the  problem of  disambiguation  would  still
remain  \textit{within}   each  such  `logical  Pauli   class',  i.e.,
different  pairs  of  ambiguous   errors  $(E_j,E_k)$  such  that  the
normalizers $E_jE_k$  correspond to the same  logical Pauli operation.
This is related  to limits imposed by the ambiguity  and can be sorted
out by  using other ASCs.  For  accessing cross terms terms  of $\chi$
across  ambiguous sets,  we use  the unitary  pre-processing described
below in Section \ref{sec:offdiag}  and \ref{sec:toggle}, based on the
method introduced by us in Ref. \cite{OSB15}.

As  an  example of  result  (\ref{eq:diag}),  for  the data  in  Table
\ref{tab:3}, the probabilities to obtain all the outcomes are
%\begin{widetext}
\begin{eqnarray}
p(++) &=& \chi_{I,I} + \chi_{Y_2,Y_2} + \chi_{X_1X_2,X_1X_2} + \chi_{X_1Z_2,X_1Z_2} 
    \nonumber \\ &+&
    2 \times [ 
                 (\textrm{Re}(\chi_{I,X_1Z_2})
     +
                 \textrm{Im}(\chi_{Y_2,X_1X_2}))\langle X_L\rangle 
    \nonumber \\ &-&
                  (\textrm{Re}(\chi_{I,Y_2})
     -
                 \textrm{Im}(\chi_{X_1X_2,X_1Z_2}))\langle Y_L\rangle
     \nonumber\\ &+&
                 (\textrm{Im}(\chi_{Y_2,X_1Z_2})
     +
                  \textrm{Re}(\chi_{I,X_1X_2}))\langle Z_L\rangle 
], \nonumber\\
p(+-) &=& \chi_{X_1,X_1} + \chi_{X_1Y_2,X_1Y_2} + \chi_{X_2,X_2} + \chi_{Z_2,Z_2} 
   \nonumber \\ &+&
    2 \times [ 
	    (\textrm{Re}(\chi_{X_1,Z_2}) - \textrm{Im}(\chi_{X_1Y_2,X_2}))\langle X_L\rangle 
    \nonumber \\
    &+& 
            (\textrm{Re}(\chi_{X_1,X_1Y_2}) - \textrm{Im}(\chi_{X_2,Z_2}))\langle Y_L\rangle \nonumber \\
     &+& (\textrm{Im}(\chi_{X_1Y_1,Z_2}) + \textrm{Re}(\chi_{X_1,X_2}))\langle Z_L\rangle 
], \nonumber \\
p(-+) &=& \chi_{Y_1,Y_1} + \chi_{Y_1Y_2,Y_1Y_2} + \chi_{Z_1X_2,Z_1X_2} + \chi_{Z_1Z_2,Z_1Z_2} \nonumber \\ &+&
    2 \times [ 
            (\textrm{Re}(\chi_{Y_1,Z_1Z_2}) + \textrm{Im}(\chi_{Y_1Y_2,Z_1X_2}))\langle X_L\rangle  \nonumber \\
    &+&  (\textrm{Im}(\chi_{Z_1X_2,Z_1Z_2})
- \textrm{Re}(\chi_{Y_1,Y_1Y_2}))\langle Y_L\rangle \nonumber \\ &+&
         (\textrm{Im}(\chi_{Y_1Y_2,Z_1Z_2}) -
            \textrm{Re}(\chi_{Y_1,Z_1X_2}))\langle Z_L\rangle],\nonumber \\
p(--) &=& \chi_{Z_1,Z_1} + \chi_{Z_1Y_2,Z_1Y_2} + \chi_{Y_1X_2,Y_1X_2} + \chi_{Y_1Z_2,Y_1Z_2} \nonumber\\ &+&
    2 \times [(\textrm{Im}(\chi_{Z_1Y_2,Y_1X_2}) - \textrm{Re}(\chi_{Z_1,Y_1Z_2})\langle X_L\rangle 
    \nonumber \\
    &+& (\textrm{Re}(\chi_{Z_1,Z_1Y_2}) 
	- \textrm{Im}(\chi_{Y_1X_2,Y_1Z_2})\langle Y_L\rangle \nonumber \\ &+&
	+ (\textrm{Im}(\chi_{Z_1Y_2,Y_1Z_2}) \textrm{Re}(\chi_{Z_1,Y_1X_2}))\langle Z_L\rangle].
\label{eq:example}
\end{eqnarray}
%\end{widetext}
By choosing input $|0\rangle_L$, one finds $p(++) = 
\chi_{I,I} + \chi_{Y_2,Y_2} + \chi_{X_1X_2,X_1X_2} + \chi_{X_1Z_2,X_1Z_2} +
2\textrm{Re}(\chi_{I,X_1X_2}) + 2\textrm{Im}(\chi_{Y_2,X_1Z_2})
\equiv C + 2\textrm{Re}(\chi_{I,X_1X_2}) + 2\textrm{Im}(\chi_{Y_2,X_1Z_2})$.
By choosing input $|{+}\rangle_L
\equiv \frac{1}{\sqrt{2}}(|0\rangle_L+|1\rangle_L)$, 
one finds $p(++) = 
C + 2\textrm{Re}(\chi_{I,X_1Z_2}) + 2\textrm{Im}(\chi_{Y_2,X_1X_2})$.
By choosing input $|{\uparrow}\rangle_L
\equiv \frac{1}{\sqrt{2}}(|0\rangle_L+i|1\rangle_L)$, 
one finds $P(++) = 
C + 2\textrm{Re}(\chi_{I,Y_2}) - 2\textrm{Im}(\chi_{X_1X_2,X_1Z_2})$.
We thus have four unknowns, given by $C$ (the diagonal
contributions), and the coefficients of $\langle X_L\rangle$,
$\langle Y_L\rangle$ and $\langle Z_L\rangle$.
One more input, say $\cos(\theta)|0_L\rangle + 
\sin(\theta)|1_L\rangle$ will suffice to determine these
4 quantities. It will thus suffice to determine $C$. More generally,
$4^k$ (the number of logical Pauli operations) preparations
are needed to solve for $C$.
When $C$ is extracted for each outcome,
then each  code gives $D^2/\gamma=2^{n-k}$ equations.   
Note that we have ignored the off-diagonal terms for ambiguous errors,
since they will be dealt with in other ASCs, where they
correspond to off-diagonal terms that are unambiguous.
 
\subsection{Preprocessing with $U$
   \label{sec:offdiag}}

For  a given  ASC,  to  derive off-diagonal  terms  between errors  in
different  ambiguous sets,  we  preprocess the  system  by applying  a
suitable   unitary  $U$,   based  on   the  idea   we   introduced  in
Ref. \cite{OSB15}.   However, even this  may allow one to  access only
the real or  imaginary part of these terms. To  access the other part,
one uses a further pre-processing described in the next Subsection.

The unitary $U$ will be in one  of two forms.  In the first form, $U =
\frac{1}{\sqrt{2}}\left(E_a    +   E_b\right)$,    in    case   $[E_a,
  E_b]\ne0$. In  the second  form, $U =  \frac{1}{\sqrt{2}}\left(E_a +
iE_b\right)$, in case $[E_a, E_b]=0$. We require $E_a$ and $E_b$ to be
mutually unambiguous in the  given ASC because otherwise, as explained
later, we obtain a situation similar to not using $U$, as far as noise
characterization is concerned.

Let  us consider  the  first case.   Suppose  the pre-processed  noisy
logical state produces an  ambiguous outcome $E_j$.  Let $g_{A_j}E_j =
E_a E_{\alpha_j}$, where  the $E_{\alpha_j}$'s constitute an ambiguous
set, and $g_{A_j} \in \{\pm  1, \pm i\}$ is the \textit{Pauli factor}.
Similarly,    let   $g_{B_j}E_j    =   E_bE_{\beta_j}$,    where   the
$E_{\beta_j}$'s constitute an ambiguous set, and $g_{B_j} \in \{\pm 1,
\pm i\}$ is a Pauli factor.

When $U(a,b)$  is applied to the  noisy logical state, and  an outcome
$x$ has been observed, then one  of the $E_j$ must have been detected,
and thus the only contributing  terms of $\mathcal{E}(\rho_L)$ will be
those      restricted      to     $|\psi_L^{\alpha_j}\rangle$      and
$|\psi_L^{\beta_j}\rangle$.   Denoting    by   $\Pi_\mathcal{C}$   the
projector to the code space  $\mathcal{C}$ of the ASC, the probability
to observe $x$ when $U(a,b)$ has been applied is:
\begin{eqnarray}
\xi(a,b,x) \equiv \textrm{Tr}\left(U 
\left[\mathcal{E}
    (|\psi_L\rangle\langle\psi_L|)\right]U^\dag
   \left(E_x\Pi_\mathcal{C}E_x\right) \right).
\label{eq:xiabx}
\end{eqnarray}
The terms within the square bracket in Eq. (\ref{eq:xiabx}) that would
make a contribution to the probability of obtaining 
ambiguous outcome $E_x$ are:
\begin{eqnarray} 
\cdots &+&
\chi_{\alpha_1,\alpha_1}|\psi_L^{(\alpha_1)}\rangle\langle\psi_L^{(\alpha_1)}| +
\chi_{\alpha_1,\alpha_2}|\psi_L^{(\alpha_1)}\rangle\langle\psi_L^{(\alpha_2)}| \nonumber \\
&+&
\chi_{\alpha_2,\alpha_1}|\psi_L^{(\alpha_2)}\rangle\langle\psi_L^{(\alpha_1)}| +
\chi_{\alpha_2,\alpha_2}|\psi_L^{(\alpha_2)}\rangle\langle\psi_L^{(\alpha_2)}| 
+ \cdots \nonumber 
\\ &+& \chi_{\alpha_1,\beta_1}|\psi_L^{(\alpha_1)}\rangle\langle\psi_L^{(\beta_1)}| +
\chi_{\beta_1,\alpha_1}|\psi_L^{(\beta_1)}\rangle\langle\psi_L^{(\alpha_1)}| +
\cdots \nonumber \\ &+&
 \chi_{\alpha_1,\beta_2}|\psi_L^{(\alpha_1)}\rangle\langle\psi_L^{(\beta_2)}| +
\chi_{\beta_2,\alpha_1}|\psi_L^{(\beta_2)}\rangle\langle\psi_L^{(\alpha_1)}|  \nonumber \\
&+& \cdots 
\label{eq:normz}
\end{eqnarray}
When the expression in Eq. (\ref{eq:normz}) is left- and right-multiplied
by $U(a,b)$, then the only resulting terms that contribute to the
lhs of Eq. (\ref{eq:xiabx}) are:
\begin{eqnarray} 
&&\cdots +
\chi_{\alpha_1,\alpha_1}|\psi_L^{(1)}\rangle\langle\psi_L^{(1)}| +
\chi_{\alpha_1,\alpha_2}g_{A_1}g_{A_2}^\ast|\psi_L^{(1)}\rangle\langle\psi_L^{(2)}| \nonumber \\
&+&
\chi_{\alpha_2,\alpha_1}g_{A_2}g_{A_1}^\ast|\psi_L^{(2)}\rangle\langle\psi_L^{(1)}| +
\chi_{\alpha_2,\alpha_2}|\psi_L^{(2)}\rangle\langle\psi_L^{(2)}| 
+ \cdots \nonumber 
\\ &+& \chi_{\alpha_1,\beta_1}g_{A_1}g_{B_1}^\ast|\psi_L^{(1)}\rangle\langle\psi_L^{(1)}| +
\chi_{\beta_1,\alpha_1}g_{B_1}g_{A_1}^\ast|\psi_L^{(1)}\rangle\langle\psi_L^{(1)}| +
\cdots \nonumber \\ &+&
 \chi_{\alpha_1,\beta_2}g_{A_1}g_{B_2}^\ast|\psi_L^{(1)}\rangle\langle\psi_L^{(2)}| +
\chi_{\beta_2,\alpha_1}g_{B_2}g_{A_1}^\ast|\psi_L^{(2)}\rangle\langle\psi_L^{(1)}| \nonumber \\
&+& \cdots
\label{eq:normz0}
\end{eqnarray}
The contribution of the first term in Eq. (\ref{eq:normz0}) to the probability
in Eq. (\ref{eq:xiabx}) would be:
\begin{eqnarray}
\epsilon_{\alpha_1,\alpha_1}&\equiv& \chi_{\alpha_1,\alpha_1}\sum_{j=1}^{2^k}\langle j_L^{(x)}|\psi_L^{(1)}\rangle\langle\psi_L^{(1)}|
j_L^{(x)}\rangle \nonumber \\
&=& \chi_{\alpha_1,\alpha_1},
\label{eq:alfa1alfa1}
\end{eqnarray}
since the traced  out quantity has support only  in the erroneous code
space     $E_x\mathcal{C}^\prime$     (i.e.,     the    code     space
$\mathcal{C}^\prime$  shifted by  the ambiguous  error).  Analogously,
the  contribution of  the  fourth term  in  Eq.  (\ref{eq:normz0})  to
Eq.               (\ref{eq:xiabx})               would              be
$\epsilon_{\alpha_2,\alpha_2}=\chi_{\alpha_2,\alpha_2}$.     In   like
fashion,  the  contribution  of  the  fifth and  sixth  terms  in  Eq.
(\ref{eq:normz0})     to     Eq.      (\ref{eq:xiabx})    would     be
$\epsilon_{\alpha_1,\beta_1}=\chi_{\alpha_1,\beta_1}$               and
$\epsilon_{\beta_1,\alpha_1}=\chi_{\beta_1,\alpha_1}$.

The contribution  of the second  term in Eq. (\ref{eq:normz0})  to the
probability in Eq. (\ref{eq:xiabx}) would be:
\begin{eqnarray}
\epsilon_{\alpha_1,\alpha_2}                                  &\equiv&
\chi_{\alpha_1,\alpha_2}g_{A_1}g_{A_2}^\ast    \sum_{j=1}^{2^k}\langle
j_L^{(x)}     |\psi_L^{(1)}\rangle     \langle\psi_L^{(2)}|
j_L^{(x)}\rangle  \nonumber  \\  
&=&  \chi_{\alpha_1,\alpha_2}g_{A_1}g_{A_2}^\ast \langle \psi_L^{(2)}|\psi_L^{(1)}\rangle
\nonumber \\
&=&  \chi_{\alpha_1,\alpha_2} g_{A_1}g_{A_2}^\ast \langle
N_{2,1}\rangle_L,
\label{eq:alfa1alfa2}
\end{eqnarray}
where  $N_{2,1}$  is  the  normalizer element  that  propagates  error
$E_{A_1}$  on a  logical ket  to $E_{A_2}$.   The contribution  of the
third   term  in   Eq.   (\ref{eq:normz0})   to  the   probability  in
Eq.      (\ref{eq:xiabx})      would      be,      analogously      to
Eq.  (\ref{eq:alfa1alfa2}),  namely,  $\epsilon_{\alpha_2,\alpha_1}  =
\chi_{\alpha_2,\alpha_1}          g_{A_2}g_{A_1}^\ast          \langle
N_{1,2}\rangle_L$.  In  like fashion, the contribution  of the seventh
and  eighth terms  in  Eq. (\ref{eq:normz0})  to Eq.  (\ref{eq:xiabx})
would                                                                be
$\epsilon_{\alpha_1,\beta_2}=\chi_{\alpha_1,\beta_2}g_{A_1}g_{B_2}^\ast\langle
N_{1,2}\rangle_L$                                                   and
$\epsilon_{\beta_2,\alpha_1}=\chi_{\beta_2,\alpha_1}g_{B_2}g_{A_1}^\ast\langle
N_{2,1}\rangle_L$.

Putting    together     all    these    $\epsilon_{\alpha_j,\alpha_k},
\epsilon_{\alpha_j,\beta_k}$, etc.,  terms into  Eq. (\ref{eq:xiabx}),
we obtain:
\begin{eqnarray}
\xi(a,b,x) &=&
 \frac{1}{2}\left(\sum_{j=1}^\gamma \chi_{\alpha_j,\alpha_j} + \chi_{\beta_j,\beta_j}+
   	\chi_{\alpha_j,\beta_j} + \chi_{\beta_j,\alpha_j}\right) \nonumber \\
   &+&
\sum_{j < k} 
\left[ \textrm{Re}\left(\chi_{\alpha_j,\alpha_k}g_{A_j}g_{A_k}^\ast\langle N_{j,k}\rangle_L\right) \right. \nonumber \\ 
   &&+~ \left.
       \textrm{Re}\left(\chi_{\beta_j,\beta_k}g_{B_j}g_{B_k}^\ast\langle N_{j,k}\rangle_L\right)\right] \nonumber \\
  &+& \sum_{j\ne k}\left[
       \textrm{Re}\left(\chi_{\alpha_j,\beta_k}g_{A_j}g_{B_k}^\ast\langle N_{j,k}\rangle_L\right)\right], 
\label{eq:xiabx0}
\end{eqnarray}
where the $\langle N\rangle$ terms,  being always real, can be removed
out of the argument of Re or Im.  

In constructing  $U(a,b)$, the  errors $E_a$ and  $E_b$ should  not be
mutually ambiguous.  Otherwise, the result is effectively  the same as
that direct measurement without  preprocessing using $U(a,b)$.  To see
this,   consider    an   application    of   this   method    to   Eq.
(\ref{eq:example}),         with         $U(X_1X_2,Y_2)         \equiv
\frac{1}{\sqrt{2}}(X_1X_2 + Y_2)$. We find:
\begin{equation}
\begin{array}{l|l|l}
\hline
   & \alpha_j, g_A & \beta_k, g_B \\
\hline
I & X_1X_2, 1  & Y_2, 1 \\
Y_2 & X_1Z_2, i  & I, 1 \\
X_1X_2 & I, 1  & X_1Z_2, i \\
X_1Z_2 & Y_2, i  & X_1X_2, -i\\
\hline
\end{array}
\end{equation}
From  (\ref{eq:xiabx0}),  it   follows  that  with  pre-processing  by
$U(X_1X_2,Y_2)$,  the  probability  expressions  in  the  example 
(\ref{eq:example}) are altered
altered, e.g.,
\begin{eqnarray}
p(++) &=& 2 \times [\textrm{Re}(\chi_{X_1X_2,Y_2}) + 
   \textrm{Re}(\chi_{I,X_1Z_2})]\langle X_L\rangle \nonumber \\
   &+& 2 \times [\textrm{Re}(\chi_{I,Y_2}) +
        - \textrm{Im}(\chi_{X_1X_2,X_1Z_2})]\langle Z_L\rangle \nonumber \\
   &+& 2\times[\textrm{Re}(\chi_{I,X_1X_2})
 + \textrm{Im}(\chi_{Y_2,X_1Z_2})]\langle Y_L\rangle
\nonumber\\
    &-& 2\times[\textrm{Im}(\chi_{Y_2,X_1X_2}) +
            \textrm{Im}(\chi_{I,X_1Z_2})].
\label{eq:U-same}
\end{eqnarray}
Thus,  the   ambiguous  errors  in  the   coefficients  of  normalizer
expectation  values  remain  the   same  even  though  the  particular
normalizer element changes.

Now, let $U =  \frac{1}{\sqrt{2}}(X_1+Z_1)$, where $X_1$ and $Z_1$ are
seen to be unambiguous for code (\ref{eq:3qec}). Set the outcome to be
`++'. This fixes $E_j$. Thus:
\begin{equation}
\left[\begin{array}{c}
E_j \\ \hline \hline
I \\ Y_2 \\ X_1X_2 \\ X_1Z_2\end{array}\right];~
\left[\begin{array}{c}
E_\alpha, g_A \\ \hline \hline
X_1,1 \\ X_1Y_2,1 \\ X_2,1 \\ Z_2,1\end{array}\right];~
\left[\begin{array}{c}
E_\beta, g_B \\ \hline \hline
Z_1,1 \\ Z_1Y_2,1 \\ Y_1X_2,-i \\ Y_1Z_2,-i\end{array}\right];~
\left[\begin{array}{c}
N \\ \hline \hline
I_L \\ Y_L \\ Z_L \\ -X_L\end{array}\right];~
\label{eq:Uab}
\end{equation}
The coefficient $\langle X_L\rangle$  to $\xi(X_1,Z_1,++)$ can be read
off  (\ref{eq:Uab}), using  (\ref{eq:xiabx0}), by  forming cross-terms
between  elements of  the second  and third  columns, such  that their
corresponding logical Pauli  operators multiply to $X_L$ up  to a sign
$g_A$. In the present case, this is seen to be
\begin{equation}
\langle X_L\rangle\left[\textrm{Im}(\chi_{X_1,Y_1Z_2} - \chi_{X_1Y_2,Y_1X_2})
+ \textrm{Re}(\chi_{X_2,Z_1Y_2} + \chi_{Z_2,Z_1})\right].
\label{eq:ReIm}
\end{equation}
We can thus form cross-terms between all ambiguous sets using suitable
$U$.

In  the second  case, $[E_a,E_b]  = 0$  and we  set $U  =  \frac{E_a +
  iE_b}{\sqrt{2}}$. As a result, instead of Eq. (\ref{eq:normz0}), one
gets:
\begin{eqnarray} 
&&\cdots +
\chi_{\alpha_1,\alpha_1}|\psi_L^{(1)}\rangle\langle\psi_L^{(1)}| +
\chi_{\alpha_1,\alpha_2}g_{A_1}g_{A_2}^\ast|\psi_L^{(1)}\rangle\langle\psi_L^{(2)}| \nonumber \\
&+&
\chi_{\alpha_2,\alpha_1}g_{A_2}g_{A_1}^\ast|\psi_L^{(2)}\rangle\langle\psi_L^{(1)}| +
\chi_{\alpha_2,\alpha_2}|\psi_L^{(2)}\rangle\langle\psi_L^{(2)}| 
+ \cdots \nonumber 
\\ &-& i\chi_{\alpha_1,\beta_1}g_{A_1}g_{B_1}^\ast|\psi_L^{(1)}\rangle\langle\psi_L^{(1)}| +
i\chi_{\beta_1,\alpha_1}g_{B_1}g_{A_1}^\ast|\psi_L^{(1)}\rangle\langle\psi_L^{(1)}| +
\cdots \nonumber \\ &-&
 i\chi_{\alpha_1,\beta_2}g_{A_1}g_{B_2}^\ast|\psi_L^{(1)}\rangle\langle\psi_L^{(2)}| +
i\chi_{\beta_2,\alpha_1}g_{B_2}g_{A_1}^\ast|\psi_L^{(2)}\rangle\langle\psi_L^{(1)}| \nonumber \\
&+& \cdots
\label{eq:normz1}
\end{eqnarray}
Consequently, one obtains in place of Eq. (\ref{eq:xiabx0}):
\begin{eqnarray}
\xi(a,b,x) &=&
 \frac{1}{2}\left(\sum_{j=1}^\gamma \chi_{\alpha_j,\alpha_j} + \chi_{\beta_j,\beta_j}+
   	\chi_{\alpha_j,\beta_j} + \chi_{\beta_j,\alpha_j}\right) \nonumber \\
   &+&
\sum_{j < k} 
\left[ \textrm{Re}\left(\chi_{\alpha_j,\alpha_k}g_{A_j}g_{A_k}^\ast\langle N_{j,k}\rangle_L\right) \right. \nonumber \\ 
   &&+~\left.
       \textrm{Re}\left(\chi_{\beta_j,\beta_k}g_{B_j}g_{B_k}^\ast\langle N_{j,k}\rangle_L\right) \right]\nonumber \\
  &+&  \sum_{j\ne k} \left[
       \textrm{Im}\left(\chi_{\alpha_j,\beta_k}g_{A_j}g_{B_k}^\ast\langle N_{j,k}\rangle_L\right)\right],
\label{eq:xiabx1}
\end{eqnarray}
where, like  before, the  $\langle N\rangle$  terms, which  are always
real, can  be removed out of  the argument of  Re or Im.  It  is worth
noting that  in Eqs.   (\ref{eq:xiabx0}) or (\ref{eq:xiabx1}),  in the
terms that  contain Pauli factors, the  matter of whether the  real or
imaginary  part  of   the  process  element  of   the  process  matrix
contributes to the measured probability,  depends on whether the Pauli
factors are of same type (real/imaginary).

\subsection{Toggling \label{sec:toggle}}

The  method  of  Section  \ref{sec:offdiag} gives  only  the  real  or
imaginary  parts of  the cross-terms.   Using an  idea we  proposed in
\cite{OSB15}, we solve this problem  by pre-processing the noisy state
even  before  applying  $U$.   Consider a  density  operator  $\rho  =
\left(  \begin{array}{cc} a  &  b \\  b^\ast &  1-a\end{array}\right)$
  subjected  to the  phase operator  given by  the diagonal  $T \equiv
  e^{i\theta_0}|0\rangle\langle0|  + e^{i\theta_1}|1\rangle\langle1|$.
  Then,  if  $\theta_0=-\theta_1=\pi/4$,  one finds  $T\rho  T^\dag  =
  \left(    \begin{array}{cc}     a    &    ib    \\     -ib^\ast    &
    1-a\end{array}\right)$, meaning that the  imaginary and real parts
    of  the off-diagonal  terms  have been  interchanged or  `toggled'
    (apart from a possible sign change).

Similarly, now we construct 
\begin{equation}
T  \equiv
\sum_{m=0}^{\sigma-1} e^{i\theta_m}\Pi_L^m,
\label{eq:toggle}
\end{equation}
where $\sigma$ is  the number of ambiguous sets  (order of ambiguity),
$\Pi_L^m$ is  the projector  to the erroneous  logical space  given by
$E_m\mathcal{C}^\prime$  ($E_m$  being  any  one error  from  each
ambiguous set),  and $\theta_m \in \{\pm  \frac{\pi}{4}\}$, with equal
entries with both signs.  Prior to  $U$, we apply the operation $T^+ =
T \oplus \mathbb{I}^\prime$,  where $\mathbb{I}^\prime$ acts trivially
outside    the    correctable   space,    i.e.,    the   code    space
$\mathcal{C}^\prime$ plus the erroneous code spaces.

For   example,  suppose   we  construct   the  toggler   $T^+$  having
$\theta_{E_\alpha}  =   -\theta_{E_\beta}=\pm\frac{\pi}{4}$,  then  in
place of (\ref{eq:Uab}) we have:
\begin{equation}
\langle X_L\rangle\left[\textrm{Re}(\chi_{X_1,Y_1Z_2} - \chi_{X_1Y_2,Y_1X_2})
+ \textrm{Im}(\chi_{X_2,Z_1Y_2} + \chi_{Z_2,Z_1})\right],
\label{eq:ReIm0}
\end{equation}
i.e.,  cross-term  $\chi_{\mu,\nu}$,  where   $\mu$  and  $\nu$  come,
respectively, from  ambiguous set $E_\alpha$ and  $E_\beta$, get their
real and imaginary parts toggled. 

The tools described in this and the preceding two subsections, as well
as  the different  ASCs, form  our repertoire  for  characterizing the
noise in the method of ASCs.

\section{Resources\label{sec:res}}

We may begin  by supposing that data from  $\gamma$ ASCs will suffice,
giving the required $D^2$ equations  to solve for the $D^2$ variables.
These $D^2$ equations will  correspond to an adjacency matrix, wherein
the  $D^2/\gamma$  rows corresponding  to  each  code  will sum  to  a
\textit{unit row}, i.e., one with  1's in all columns.  Thus there are
(at least)  $\gamma-1$ constraints  among the $D^2$  equations. Adding
one  more code  will  introduce $D^2/\gamma$  equations  and one  more
constraint  i.e.,  $2^{n-k}-1$ constraints.   If  there  are no  other
constraints in  the first $D^2$ rows,  and if $2^{n-k}-1\geq\gamma-1$,
i.e., $n-k\geq  p$, then  the remaining required  linearly independent
equations can  be found  from the last  code.  Thus, in  general, with
$\gamma$-fold full degeneracy, the necessary number of preparations is
$\gamma+1$.

More  generally,  because  of  the  failure of  QEC  isomorphism  with
ambiguous codes,  of the $O\left(4^m  \times 4^m\right)$ terms  in the
process     matrix,     only     $O\left(\frac{4^m}{\gamma}     \times
\frac{4^m}{\gamma}\right)$  independent terms  can  be determined  per
ASC,   implying   that   a   full   characterization   would   require
$\mu=O(\gamma^2)$  different ASC's.   Also,  syndrome measurements  on
each  ASC yields  $D^2/\gamma  = 4^m/\gamma$  outcomes.   We may  thus
estimate  that   the  number  of  configurations  required   is  $c  =
O\left(\frac{16^m}{\gamma^2}/\frac{4^m}{\gamma}\right)                =
O\left(\frac{4^m}{\gamma}\right)$ per ASC.  Thus in all, counting each
ASC   as  a   separate   configuration,  we   require  $\mu\times   c$
configurations, i.e.,  $O(\gamma4^m)$, meaning that there  is a factor
$\gamma$ excess when using ambiguous codes.  (Moreover each code would
require up to $4^k$ state preparations for disambiguation of the Pauli
logical classes.)   This can be considered  as a time cost  to pay for
the saving  in `space', i.e., in  terms of number  of entangled qubits
used.

Now we present  an example of applying QASCD,  with three 4-qubit ASCs
being used to characterize a 2-qubit noise.

\section{Illustration using a family of three 4-qubit ambiguous codes
\label{sec:example}}

Consider the  $[[4,1]]$ ASC  ${\bf C}^1$ for  arbitrary errors  on the
first  two qubits,  constructed  by  dropping the  last  qubit of  the
$[[5,1]]$ QECC of Ref. \cite{LMP+96}:
\begin{eqnarray}
|0_L^1\rangle_4&=&\frac{1}{2\sqrt{2}} \left(-|0000\rangle + |0010\rangle +
|0101\rangle + |0111\rangle \right.\nonumber  \\
&&~~~~~~~\left.-|1001\rangle+|1011\rangle+|1100\rangle+|1110\rangle\right) \nonumber  \\
|1_L^1\rangle_4&=&\frac{1}{2\sqrt{2}} \left(-|1111\rangle + |1101\rangle +
|1010\rangle + |1000\rangle \right.\nonumber  \\
&&~~~~\left.-|0110\rangle+|0100\rangle+|0011\rangle+|0001\rangle\right),\\\nonumber
\label{eq:41}
\end{eqnarray}
whose stabilizer generators are $XIIX, YIXY$ and $YYZZ$.
The following equation  presents two other such codes ${\bf C}^2$ and 
${\bf C}^{3}$ which are two fold amiguous:
\begin{eqnarray}
\label{eq:cw}
|0^2_L\rangle_4  &=&  H_{ZY}^{\otimes4}|0_L\rangle,~
|1^2_L\rangle_4=H_{ZY}^{\otimes4}|1_L\rangle_4, \nonumber \\
|0^{3}_L\rangle_4  &=& H_{YX}^{\otimes4}|0_L\rangle_4,
|1^{3}_L\rangle_4=H_{YX}^{\otimes4}|1_L\rangle_4,
\end{eqnarray}
 where $H_{ZY}=\frac{1}{\sqrt{2}}
(|0\rangle\langle0|  +  i|0\rangle\langle1|  +  i|1\rangle\langle0|  +
|1\rangle\langle1|)$, $H_{YX}=\frac{1}{2}((1+i)|0\rangle\langle0| + (1
+    i    )    |0\rangle\langle1|    -    (1-i)|1\rangle\langle0|    +
(1-i)|1\rangle\langle1|)$. The   corresponding   stabilizer generators   
and the error syndromes are   given   in   Table \ref{tab:syndrome}.

By method described  in Sec. \ref{sec:Xr}, the  statistics of syndrome
outcomes  on QECs  ${\bf C}^1$,  ${\bf  C}^2$ and  ${\bf C}^{3}$,  can
completely determine the process  matrix $\chi_{m,n}$ corresponding to
an arbitrary 2-qubit noise $\mathcal{E}$.   It can be noticed from the
Table \ref{tab:syndrome},  that the normalizer corresponds  to logical
$Y_L$. By  the direct measurement,  as in Eq. (\ref{eq:diag}),  we get
$\chi_{\alpha_1,\alpha_1}+\chi_{\alpha_2,\alpha_2}         +         2
\textrm{Re}\left(\chi_{\alpha_1,\alpha_2}\langle   Y_L\rangle\right)$.
By choosing  any state of  ${\bf C}^1$ other  than $|\uparrow\rangle$,
$\langle  Y_L\rangle$ vanishes.  The direct  syndrome measurements  on
suitably prepared ${\bf C}^1$ yields the following expressions,
\begin{eqnarray}
&&\chi_{I,I}+\chi_{Y_2,Y_2}=a_1, ~\chi_{X_1,X_1}+\chi_{X_1Y_2,X_1Y_2}=b_1,\nonumber\\
&&\chi_{X_2,X_2}+\chi_{Z_2,Z_2}=c_1, \chi_{X_1X_2,X_1X_2}+\chi_{X_1Z_2,X_1Z_2}=d_1,\nonumber\\
&&\chi_{Y_1X_2,Y_1X_2}+\chi_{Y_1Z_2,Y_1Z_2}=e_1,~\chi_{Y_1,Y_1}+\chi_{Y_1Y_2,Y_1Y_2}=f_1,\nonumber\\
&&\chi_{Z_1Z_2,Z_1Z_2}+\chi_{Z_1Z_2,Z_1X_2}=g_1,\chi_{Z_1,Z_1}+\chi_{Z_1Y_2,Z_1Y_2}=h_1.\nonumber\\
\label{eq:diag1}
\eea   
Similarly  procedure followed on   ${\bf C}^2$  yields  
\bea
&&\chi_{I,I}+\chi_{Z_2,Z_2}=a_2,~\chi_{X_1,X_1}+\chi_{X_1Z_2,X_1Z_2}=b_2,\nonumber\\ 
&&\chi_{Y_1,Y_1}+\chi_{Y_1Z_1,Y_1Z_2}=c_2,~\chi_{Z_1,Z_1}+\chi_{Z_1Z_2,Z_1Z_2}=d_2.\nonumber\\
\label{eq:diag2}
\eea
From   ${\bf C}^3$ we obtain the following expressions
\bea
\chi_{I,I} + \chi_{X_2,X_2}  = a_3,~\chi_{X_1,X_1}+\chi_{X_1X_2,X_1X_2}=b_3,\nonumber\\
\chi_{Y_1,Y_1}+\chi_{Y_1X_2,Y_1X_2}=c_3,~\chi_{Z_1,Z_1}+\chi_{Z_1X_2,Z_1X_2}=d_3.\nonumber\\
\label{eq:diag3}
\eea

\begin{table}
\begin{tabular}{|l|l|l|l|l|l|l|l|l|l|l|l|l|l|l|l|l|}\hline
${\bf C}^1$&$II $&$X_1$&$X_2$&$Y_1$&$Z_1$&$XX$ &$YX $&$ZX$ \\ 
             &$Y_2$&$XY$ &$Z_2$&$YY$ & $ZY$&$XZ$ &$YZ$ &$ZZ$   \\ \hline
$XIIX$       & +   & +   & +   & --  & --  & +   & --  & --   \\ \hline
$YIXY$       & +   & --  &  +  & +   & --  & --  & +   & --  \\ \hline
$YYZZ$       & +   & --  & --  & +   & --  & +   & --  &  + \\ \hline
\hline
${\bf C}^{2}$&$II $&$X_1$&$X_2$&$Y_1$&$Z_1$&$XX $&$YX$ &$ZX$ \\ 
                      &$Z_2$&$XZ$ &$Y_2$&$YZ$ & $ZZ$&$XY$ &$YY$ &$ZY$   \\ \hline
$IZZX$                & +   & +   & --  & +   & +   & --  & --  & --     \\ \hline
$XIIX$                & +   & +   &  +  & --  & --  & +   & --  & --     \\ \hline
$YZYZ$                & +   & --  & --  & +   & --  & +   & --  & +     \\ \hline
\hline
${\bf C}^{3}$&$II $&$X_1$&$Y_1$&$Y_2$&$Z_1$&$XY $&$YY $&$ZY$ \\ 
                            &$X_2$&$XX$ &$YX $&$Z_2$& $ZX$&$XZ$ &$YZ$ &$ZZ$   \\ \hline
$IXXZ$                      & +   & +   & +   &--   & +   & --  & --  & --  \\ \hline
$XIXZ$                      & +   & +   & --  & +   & --  & +   & --  & -- \\ \hline
$YXYX$                      & +   & --  & +   & --  & --  & +   & --  & + \\ \hline
\end{tabular} 
\caption{  Ambiguous class  for  the three  4-qubit codes.  The
    Hadamard operation $H_{ZY}$ ($H_{YX}$)  toggles errors $Z$ and $Y$
    (errors $Y$ and $X$) while keeping  error $X$ ($Z$) fixed, and the
    above  syndromes  are  corresponding   toggled  versions  of  each
    other.}
\label{tab:syndrome}
\end{table}

The above 16 expressions suffice  to determine the diagonal terms
of the process matrix.
To  demonstrate  how  the  method works  for  off-diagonal  terms,  we
consider its application to the  noise 
\bea \mathcal {E}_A(\rho_L) &&=
\delta   \rho_L    +   \frac{1-\delta}{5}\left(X_1   \rho_L    X_1   +
XZ\rho_LXZ+Y_2\rho_LY_2\right.\nonumber\\ &&\left. +X_2\rho_LX_2+XX\rho_L
XX\right)+\frac{1}{6}\left(                          (a+ib)X1\rho_LX_2
\right.\nonumber\\   &&+(c+id)\rho_L  XX   +   (e+if)XZ\rho_L  Y_2   +
\textrm{c.c})  
\eea  

In   the  present  case,  for   solving  the off-diagonal terms  using 
Eq. (\ref{eq:xiabx1}), the following  set of  linearly independent
equations for  off-diagonal terms  are obtained by  performing 
unitary   operations   $U(a,b)$    followed   by  syndrome
measurements on ${\bf C}^1$, ${\bf C}^2$ and ${\bf C}^3$
respectively
\bea
\xi(I,X_1X_2,I)&=&\chi_{I,I}+\chi_{X_1X_2,X_1X_2}+\chi_{Y_2,Y_2}+\chi_{X_1Z_2,X_1Z_2}\nonumber\\
 &&+\textrm{Im}(I,X_1X_2)+\textrm{Re}(Y_2,X_1Z_2),\nonumber\\
\xi(I,X_1X_2,X_1)&=&\chi_{X_1,X_1}+\chi_{X_2,X_2}+\chi_{Y_2,Y_2}+\chi_{X_1Z_2,X_1Z_2}\nonumber\\
&&+\textrm{Im}(X_1,X_2)-\textrm{Re}(Y_2,X_1Z_2),\nonumber\\
\xi(I,X_1X_2,I)&=&\chi_{I,I}+\chi_{X_1X_2,X_1X_2}+\chi_{X_1,X_1}+\chi_{X_2,X_2}\nonumber\\
&&+\textrm{Im}(I,X_1X_2)-\textrm{Im}(X_1,X_2).\nonumber\\
\label{eq:example0} 
\eea
In  Eq. (\ref{eq:example0}),  the
diagonal  terms   are  obtained  without pre-processing with 
unitaries using in Eq. (\ref{eq:diag1}), (\ref{eq:diag2}) and (\ref{eq:diag3}). 
Solving the  above set  of equations we obtain the off-diagonal terms of 
the process matrix corresponding to $\mathcal{E}_A$ :                  
\bea
\textrm{Im}(I,X_1X_2)&=&\frac{1}{2}(\mathcal{O}_1+\mathcal{O}_2+\mathcal{O}_3)=\frac{c}{6},\nonumber\\ 
\textrm{Im}(X_1,X_2)&=&\frac{1}{2}(\mathcal{O}_1+\mathcal{O}_2-\mathcal{O}_3)=\frac{a}{6},\nonumber\\ 
\textrm{Re}(Y_2,X_1Z_2)&=&\mathcal{O}_1-\frac{1}{2}(\mathcal{O}_1+\mathcal{O}_2+\mathcal{O}_3)=\frac{f}{6},
\eea
where $\mathcal{O}_1=\xi(I,X_1X_2,I)-(\chi_{I,I}+\chi_{X_1X_2,X_1X_2}+\chi_{Y_2,Y_2}+\chi_{X_1Z_2,X_1Z_2})$,
$\mathcal{O}_2=\xi(I,X_1X_2,X_1)-(\chi_{X_1,X_1}+\chi_{X_2,X_2}+\chi_{Y_2,Y_2}+\chi_{X_1Z_2,X_1Z_2})$
and $\mathcal{O}_3=\xi(I,X_1X_2,I)-(\chi_{I,I}+\chi_{X_1X_2,X_1X_2}+\chi_{X_1,X_1}+\chi_{X_2,X_2})$.

The   real    or   imaginary    counterparts   of    the   expressions
Eq. (\ref{eq:example}) are obtained  by preprocessing the noisy states
with   the    corresponding   toggling   operations.    For   code
$\textbf{C}^1$, note that $I$ and $Y_2$ are ambiguous, and thus cannot
have different toggler signs. On the  other hand, we want them both to
have different  toggler signs  than $X_1X_2$  and $X_1Z_2$,  which are
also ambiguous. Thus, one required toggling operation would be:
\begin{eqnarray}
T^+_j&=&\frac{1+i}{\sqrt{2}}\left(\Pi_{{\bf C}^1} +
   X_1\Pi_{{\bf C}^1}X_1+ 
%\right.\nonumber\\ &&\left. 
Y_1\Pi_{{\bf C}^1}Y_1+
Z_1\Pi_{{\bf C}^1}Z_1\right)\nonumber\\
&+&\frac{1-i}{\sqrt{2}}
\left(X_1X_2\Pi_{{\bf C}^1}X_1X_2+ 
X_2\Pi_{{\bf C}^1}X_2  \right.\nonumber\\
&&\left. +Y_1X_2\Pi_{{\bf C}^1}Y_1X_2+
Z_1X_2\Pi_{{\bf C}^1}Z_1X_2\right),
\end{eqnarray}
and similarly for the codes  ${\bf C}^{j}$ ($j \in \{2,3\})$.  The
expressions obtained by pre-processing the noisy ASCs with unitary and
toggling are
\begin{eqnarray}
\xi^\prime(I,X_1X_2,I)&=&\chi_{I,I}+\chi_{X_1X_2,X_1X_2}+\chi_{Y_2,Y_2}+\chi_{X_1Z_2,X_1Z_2}\nonumber\\
 &&+\textrm{Re}(I,X_1X_2)+\textrm{Im}(Y_2,X_1Z_2),\nonumber\\
\xi^\prime(I,X_1X_2,X_1)&=&\chi_{X_1,X_1}+\chi_{X_2,X_2}+\chi_{Y_2,Y_2}+\chi_{X_1Z_2,X_1Z_2}\nonumber\\
&&+\textrm{Re}(X_1,X_2)+\textrm{Im}(Y_2,X_1Z_2),\nonumber\\
\xi^\prime(I,X_1X_2,I)&=&\chi_{I,I}+\chi_{X_1X_2,X_1X_2}+\chi_{X_1,X_1}+\chi_{X_2,X_2}\nonumber\\
&&+\textrm{Re}(I,X_1X_2)+\textrm{Re}(X_1,X_2).\nonumber\\
\label{eq:example1} 
\eea
Solving the above set of equations we have the real or imaginary parts 
of the off-diagonal terms of the process matrix that wre undetermined by Eq. (\ref{eq:example}) 
with out toggling: 
\bea
\textrm{Re}(I,X_1X_2)&=&\frac{1}{2}(\mathcal{O}_1^\prime-\mathcal{O}_2^\prime+\mathcal{O}_3^\prime)=\frac{c}{6},\nonumber\\ 
\textrm{Re}(X_1,X_2)&=&\frac{1}{2}(-\mathcal{O}_1^\prime+\mathcal{O}_2^\prime+\mathcal{O}_3^\prime)=\frac{a}{6},\nonumber\\ 
\textrm{Im}(Y_2,X_1Z_2)&=&\frac{1}{2}(\mathcal{O}_1^\prime+\mathcal{O}_2^\prime-\mathcal{O}_3^\prime)=\frac{f}{6},
\eea
where
$\mathcal{O}_1^\prime=\xi^\prime(I,X_1X_2,I)-(\chi_{I,I}+\chi_{X_1X_2,X_1X_2}+\chi_{Y_2,Y_2}+\chi_{X_1Z_2,X_1Z_2})$
$\mathcal{O}_2^\prime=\xi^\prime(I,X_1X_2,X_1)-(\chi_{X_1,X_1}+\chi_{X_2,X_2}+\chi_{Y_2,Y_2}+\chi_{X_1Z_2,X_1Z_2})$
$\mathcal{O}_3^\prime= \xi^\prime(I,X_1X_2,I)-(\chi_{I,I}+\chi_{X_1X_2,X_1X_2}+\chi_{X_1,X_1}+\chi_{X_2,X_2})$.

\section{Discussion and conclusion\label{sec:conclu}}

We developed the concept of  ambiguous stabilizer codes, which exploit
the  stabilizer formalism  for quantum  error characterization  rather
than for quantum error  correction.  We presented different procedures
for constructing  an $[[n,k]]$ ASC that  ambiguously detects arbitrary
errors on  $m$ known qubit coordinates  ($m < n$). The  Pauli operator
basis  for  this  set  of  errors  forms a  group.   The  ASC  can  be
characterized as a  quotient group $\mathcal{P}_m/\mathfrak{B}$, where
$\mathfrak{B}$ is the set of $m$-qubit Pauli errors ambiguous with the
no-error syndrome.  The cosets  of $\mathfrak{B}$ form other ambiguous
sets of errors.

ASCs cannot be used for quantum  error correction, except if the basis
elements of  the noise is  known to have at  most a single  element in
each of  the ambiguous sets  of the  ASC. Quite generally,  a suitable
collection of ASCs can be  employed for characterizing noise, and this
is the chief  application of ASCs. The  code length for an  ASC can be
smaller than demanded  by the requirement of  error correction, making
state   preparations   potentially   simpler  from   an   experimental
perspective than  for the  techniques of Refs.   \cite{ML06,OSB15}. We
developed   a  protocol,   ``quantum  ASC-based   characterization  of
dynamics'' (QASCD),  for this purpose,  which, in comparison  with the
use of  conventional stabilizer  codes for CQD  \cite{OSB15}, requires
smaller  code length,  but  at  the cost  of  more  number of  quantum
operations and  classical post-processing.  We illustrated  our method
using  an example  of characterization  of a  toy 2-qubit  noise using
three 4-qubit ASCs.

\acknowledgments

OS and RS acknowledge financial  support through DST, Govt.  of India,
for the  project SR/S2/LOP-02/2012.  OS acknowledges  academic support
from the Manipal University graduate program.

\bibliography{QECPT}

\end{document}